# Energy resolution of a photon-counting silicon strip detector

Erik Fredenberg[*,a], Mats Lundqvist[b], Björn Cederström[a], Magnus Åslund[b], Mats Danielsson[a]

[a] Department of Physics, Royal Institute of Technology, AlbaNova University Center, SE-106 91 Stockholm, Sweden
[b] Sectra Mamea AB, Smidesvägen 5, SE-171 41 Solna, Sweden

**Abstract**

A photon-counting silicon strip detector with two energy thresholds was investigated for spectral x-ray imaging in a mammography system. Preliminary studies already indicate clinical benefit of the detector, and the purpose of the present study is optimization with respect to energy resolution. Factors relevant for the energy response were measured, simulated, or gathered from previous studies, and used as input parameters to a cascaded detector model. Threshold scans over several x-ray spectra were used to calibrate threshold levels to energy, and to validate the model. The energy resolution of the detector assembly was assessed to range over $\Delta E/E = 0.12$ to $0.26$ in the mammography region. Electronic noise dominated the peak broadening, followed by charge sharing between adjacent detector strips, and a channel-to-channel threshold spread. The energy resolution may be improved substantially if these effects are reduced to a minimum. Anti-coincidence logic mitigated double counting from charge sharing, but erased the energy resolution of all detected events, and optimization of the logic is desirable. Pile-up was found to be of minor importance at typical mammography rates.

## 1. Introduction

X-ray mammography is an effective and wide-spread method to diagnose breast cancer, but it is also technically demanding [1]. Two major challenges that face the modality are the small signal differences between lesions and breast tissue, and the lumpy backgrounds that are caused by superposition of glandular structures.

Spectral imaging is a method to extract information about the object constituents by the material specific energy dependence of x-ray attenuation [2, 3]. In mammography, there are at least two potential benefits of this approach compared to non-energy resolved imaging. (1) The signal-to-quantum-noise ratio may be optimized with respect to its energy dependence; photons at energies with larger agent-to-background contrast can be assigned a greater weight [4, 5]. (2) The signal-to-background-noise ratio can be optimized by minimization of the background clutter contrast. A weighted subtraction of two images acquired at different mean energies cancels the contrast between any two materials (adipose and glandular tissue) whereas all other materials (lesions) to some degree remain visible. The contrast in the subtracted image is greatly improved if the lesion is enhanced by a contrast agent with an absorption edge in the energy interval, which provides a large difference in attenuation [6, 7, 8, 9, 10].

One way of obtaining spectral information is to use two or more input spectra. For imaging with clinical x-ray sources, this most often translates into several exposures with different beam qualities (different acceleration voltages, filtering, and anode materials) [6, 7, 8]. Results of the dual-spectra approach are promising, but the examination may be lengthy with increased risk of motion blur and discomfort for the patient. This may be solved by instead using an energy sensitive detector, which has been pursued with sandwich detectors [11, 12]. For both of the above approaches, however, the effectiveness may be impaired due to overlap of the spectra, and a limited flexibility in choice of spectra and energy levels. In recent years, photon-counting silicon detectors with high intrinsic energy resolution and, in principle, an unlimited number of energy levels (electronic spectrum-splitting) have been introduced as another option [9, 10, 13, 14].

An objective of the EU-funded HighReX project is to investigate the benefits of spectral imaging in mammography [15]. The systems used in the HighReX project are based on the Sectra MicroDose Mammography (MDM) system,[1] which is a scanning multi-slit full-field digital mammography system with a photon-counting silicon strip detector [16, 17, 18]. An advantage of this geometry in a spectral imaging context is efficient intrinsic scatter rejection [19, 20, 21].

We have investigated the energy response of a proto-type detector for the HighReX project on a system level. The major factors that affect the energy response have been identified, and used as input to a cascaded detector model. The purpose of the model is detector optimization with respect to energy resolution. Optimal energy resolution will improve performance when the detector is employed for spectral imaging within the HighReX project, in particular when using a K-edge contrast agent such as iodine. Knowledge of the energy resolution will also be essential for simulating contrast- and noncontrast-enhanced spectral imaging with the detector.

---

[*]Corresponding author
   *Email address:* fberg@mi.physics.kth.se (Erik Fredenberg)
   *URL:* http://www.mi.physics.kth.se (Erik Fredenberg)

[1]Sectra Mamea AB, Solna, Sweden

## 2. Materials and Methods

### 2.1. Description of the system and detector

Because the systems used in the HighReX project are modifications of the MDM system, and because an MDM system was used for testing the detector, we abridge our discussion to consider only the MDM system. It comprises a tungsten target x-ray tube with aluminum filtration, a pre-collimator, and an image receptor, all mounted on a common arm (Fig. 1). The image receptor consists of several modules of silicon strip detectors with corresponding collimator slits in the pre-breast collimator. To acquire an image, the arm is rotated around the center of the source so that the detector modules and pre-collimator are scanned across the object. In Fig. 1 and henceforth, $x$ refers to the detector strip direction and $y$ to the scan direction.

The detector modules were fabricated on 500 $\mu$m thick n-type silicon wafers with p-doped strips at a pitch of 50 $\mu$m. Each strip thus forms a separate PIN-diode, which is depleted by a 150 V bias voltage. Aluminum strands are DC-coupled to the strips, and wire bonded to the read-out electronics. To obtain high quantum efficiency despite the relatively low atomic number of silicon, the modules are arranged edge-on to the x-ray beam [13]. Interactions in the guard ring are avoided by irradiating the detector modules at a slight angle [17], which yields an effective thickness of approximately 4 mm. Scatter shields between the modules block detector-to-detector scatter. The silicon detector modules are in many ways similar to the ones that are used in the MDM system [18, 17], whereas the read-out electronics to a larger degree are different from previous versions [17, 22].

Each detector strip is connected to a preamplifier and shaper, which are fast enough to allow single photon counting. The pulse height depends on the released charge in the silicon, and thus on the energy of the impinging photon. An average of 273 electron-hole pairs are created for each keV photon energy, whereas the equivalent noise charge is in the order of a few hundred electrons, and a low-energy threshold at a few keV in a discriminator following the shaper ensures that the electronic noise does not affect the number of detected counts. All remaining pulses are sorted into two energy bins by an additional high-energy threshold, and registered by two counters. A preamplifier with discriminator and counters are referred to as a channel, and all channels are implemented in an application specific integrated circuit (ASIC). The gain of the preamplifier varies slightly between the channels, and to compensate for this, the threshold levels of individual channels were trimmed in 3 bits towards either the electronic noise floor or some discontinuity in the input spectrum. On-chip current-based 8-bit digital-to-analog converters define the global high- and low-energy threshold levels.

Charge sharing between adjacent detector strips may increase image noise at low spatial frequencies and degrade the spatial resolution if the charge is large enough to be registered by both channels (double counting). The energy resolution is also affected because all charge is not collected into a single pulse. The present ASIC implements anti-coincidence (AC) logic, which distinguishes charge-shared events by a simultaneous detection of pulses that reach over the low-energy threshold in adjacent channels. The first detected pulse, which is generally the largest one, increments the high-energy bin, whereas the slower pulse is rejected. Double counting is thus avoided, which improves spatial resolution and noise, but all energy information is lost. The AC logic cannot be turned off in the present ASIC, but is disabled by masking every other channel. This procedure reduces the efficiency and is an option for physical evaluation only, not for clinical imaging.

### 2.2. Modeling the detector

The energy response function of the detector was modeled using the MATLAB software package[2] as a semi-empirical cascade of several detector effects. These were grouped into 8 categories, which are outlined in the bottom part of Fig. 1 and described in detail below. Some of the steps in the cascade require measured input parameters, and the procedures to find these are described in the next section.

(1) Quantum efficiency was calculated with published linear absorption coefficients [23]. Charge collection on the aluminum strands was assumed ideal so that the full energy deposition of photo-electric events was detected. The low-energy threshold was assumed to reject the detection of Compton scattered events so that scattering only contributed to filtering of the beam. Secondary photo-electric interactions of scattered photons in adjacent detector modules was eliminated by scatter shields, and secondary interactions in the detector module of the first interaction was ignored because of the large angular spread of Compton scattering. Rayleigh scattering was excluded altogether because of a relatively small cross section at hard x-ray energies. Fluorescence is generally a minor problem in silicon detectors at hard x-ray energies [24], and it was ignored in the model. Elaborate motivations for ignoring

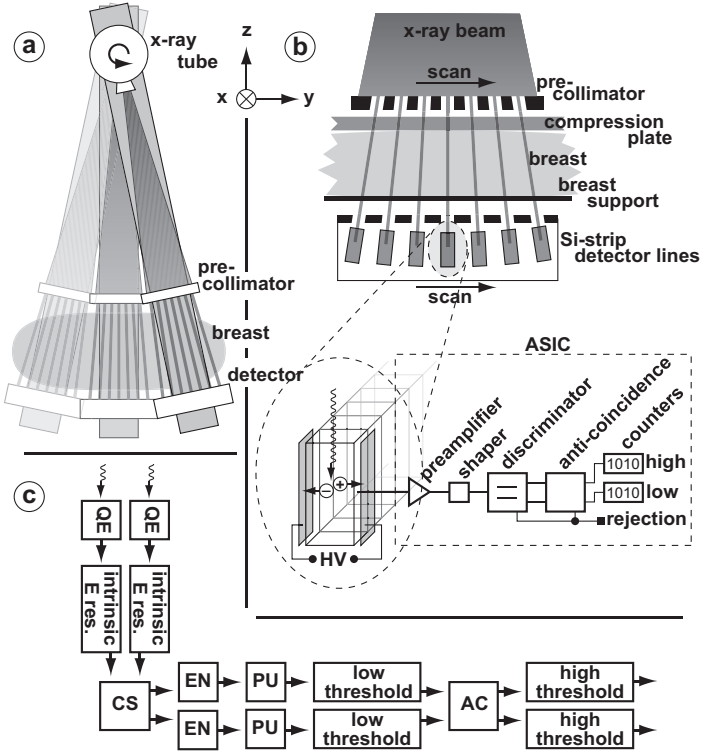

Figure 1: **a:** In the MDM system, the arm is rotated around the center of the source to acquire an image. **b:** Closeup of the detector assembly and the electronics. **c:** Block diagram of the cascaded detector model for two adjacent channels. The model includes: (1) Quantum efficiency (QE). (2) Intrinsic lower energy resolution of silicon. (3) Charge sharing (CS). (4) Electronic noise (EN). (5) Pile-up in the shaper (PU). (6) Nonlinearity of the shaper and thresholds, and bit resolution. (7) Channel-to-channel spread of the thresholds. (8) Anti-coincidence logic (AC) with leakage and chance coincidence (CC).

scattering and fluorescence can be found at the end of this section.

(2) With a relatively large number of released charge pairs at each photon conversion, silicon has good intrinsic energy resolution. The peak was modeled as normal distributed with standard deviation (in units of eV) $\sigma_i = \sqrt{E\eta\epsilon}$, where $\epsilon = 3.66$ eV is the mean energy needed to create an electron-hole pair in silicon, and $\eta = 0.115$ is the Fano factor for silicon [24]. The full-width-at-half-maximum (FWHM) of the peak is 0.2–0.3 keV in the interval 20–40 keV.

(3) Charge sharing results in loss of detected charge and a corresponding spread towards lower pulse heights in the channel of interaction, and a reversed probability for detection of charge from interactions in adjacent strips. We used the probability distribution from a previously developed computer model to predict the effects of charge sharing [18]. With a 50 $\mu$m strip pitch, charge sharing has a relatively large impact on energy resolution with peak widths ranging from 1.8 to 1.4 keV FWHM in the 20–40 keV interval and with heavy tails towards lower energies.

(4) Electronic noise generally has a negligible effect on the number of detected events in a photon counting detector, but the energy resolution is affected. The equivalent noise charge in a similar ASIC without silicon detector attached has been found to be $\sigma_{add} = 200$ electrons r.m.s. at [17], but we can expect a higher level in this study because of the added detector capacitance and leakage current.

(5) Pile-up occurs mainly in the shaper. For typical mammography rates of $R < 500$ kHz, and shaper dead times $\tau_s < 200$ ns, the product $R\tau_s \ll 1$, and pile-up is a relatively small effect. In that case, there is no need to distinguish between paralyzable and non-paralyzable shaper behavior [24]. Ignoring multiple pile-up, the detected count-rate is then

$$r_{pu} \approx R - R^2\tau_s, \qquad (1)$$

where $R$ is the true rate without pile-up. The distribution of two piled-up pulses with partial overlap was simplified into a rect function extending from $\min(E_1, E_2)$ to $\text{sum}(E_1, E_2)$, where $E_1$ and $E_2$ are the energies of the impinging photons.

(6) The combined energy response of shaper and discriminator is approximately linear at low energies and then saturates. The nonlinearity at higher energies was found empirically to be well described by an inverse power function so that the threshold level ($T$) as a function of energy ($E$) is

$$T(E) = \begin{cases} C_1 E + C_2 & \text{for } E < C_6 \\ C_3 E^{-C_4} + C_5 & \text{for } E \geq C_6 \end{cases}, \qquad (2)$$

where the coefficients $C_1$–$C_6$ are free parameters. A reduction to only four parameters is achieved by requiring $T$ and $dT/dE$ to be continuous.

(7) Small deviations in the threshold levels of individual channels remained after trimming because of a limited bit depth and slightly different energy dependence of the channels. This resulted in an energy dependent channel-to-channel spread, which was modeled as normal distributed and increasing away from the trimming point. The spread in a single module of a similar detector has been measured to approximately 0.9 keV FWHM [17].

(8) Chance coincidence in the AC logic occurs at a rate $r_{ch} = R[1 - \exp(-2R\tau_{ac})] \approx 2R^2\tau_{ac}$, where $\tau_{ac}$ is the AC time window and the approximation is for $2R\tau_{ac} \ll 1$ [24]. The count-rates in the two bins are then

$$\begin{aligned} r_{lo} &= R_{lo} - 2r_{ch} \approx R_{lo} - 4R^2\tau_{ac}, \quad \text{and} \\ r_{hi} &= R_{hi} + r_{ch}(1 + \xi_{cc}) \approx \\ &\approx R_{hi} + 2R^2\tau_{ac}(1 + \xi_{cc}), \end{aligned} \qquad (3)$$

where $\xi_{cc}$ is the leakage of the logic. Combining Eqs. (1) and (3), the total count-rate is $r_{sum} \approx R - R^2[\tau_s + 2\tau_{ac}(1 - \xi_{cc})]$, and we note that the impact of chance coincidence is twice that of pile-up if there is no leakage. A preliminary electronics test revealed, however, that if two simultaneous pulses are similar in size, the AC logic cannot make a correct decision and both pulses are directed to the respective high-energy bins. This is a relatively unlikely situation for charge-shared events because it requires interaction close to the border between two strips. It is, however, more likely in the case of chance coincidence because the energies are higher, which results in similar-sized pulses due to the nonlinear shaper output. We can thus expect two leakage coefficients of the AC logic, $\xi_{cc} > \xi_{cs}$, for chance coincidence and charge sharing, which have to be determined separately.

Published mammography spectra [25] were used as input to the model for comparison to measurements. The energy resolution of the high-energy threshold was evaluated as $\Delta E/E$, where $\Delta E$ is the FWHM of the predicted response to a delta peak.

To verify the assumption that Compton scattered photons pose a minor problem, a simple geometrical model was set up that traced a photon through the center of a detector module. The Klein-Nishina cross section was used to calculate a probability function for scattering angle and deposited energy [26]. Accordingly, energy deposition increases with incident photon energy, and for the hardest spectrum considered in this study (40 kV and 3 mm aluminum filtration), the mean deposited energy was found to be 1.5 keV with a maximum (Compton edge) of 5.4 keV. It is hence safe to assume that scattered events are rejected in the detector strip of the primary interaction for typical low-energy threshold levels. The detected scatter-to-primary ratio for first order secondary interactions of scattered photons was 2.1%, with a maximum of 2.8% for 40 keV photons. We ignored this amount, which is similar to what may detected from scattering in an object; 2.1% was measured for a 50 mm breast at 38 kV and 0.5 mm aluminum filtration in a similar geometry [21].

It cannot be excluded that fluorescent photons escape the relatively narrow strips of the detector. Therefore, the size of the escape peak was calculated according to a previous, experimentally verified, study [27]. In summary, 92% of the absorbed photons eject a K-electron, the K-fluorescent yield of silicon is 4.3%, and the energy of the fluorescent photon is 1.74 keV. A similar geometry as for calculating scattering was used. We found that the relative intensity of the escape peak was 0.3% at 15 keV and decreasing with energy because of deeper interactions in the silicon. 15 keV is in the lowermost region of typical mammography spectra, and fluorescence can hence be confidently ignored.

### 2.3. Measurements on the detector

A complete detector assembly with a total of 89856 channels, was mounted on a standard MDM system. The low-energy threshold levels were trimmed towards the electronic noise to minimize the variance in count-rates between channels. The high-energy thresholds were trimmed against the steep derivative at the K absorption edge (33.2 keV) of an iodine filtered 40 kVp spectrum. The air kerma was monitored with an ion chamber,[3] and, knowing the exposure time, converted to flux using published spectra [25], attenuation and energy absorption coefficients [23].

Integral pulse height spectra were acquired by scanning the high- and low-energy thresholds of 144 channels over several incident energy spectra. The tungsten spectrum was filtered with a total of 3 mm aluminum to make it relatively distinct. In the following, the words "threshold scan" and "integral pulse height spectrum" are used interchangeably when relating to this procedure. The mean pulse height spectra between all channels were used to calibrate the global threshold level to energy and estimating the electronic noise by fitting the coefficients of Eq. (2) and $\sigma_{add}$ for each energy bin separately, keeping all other model parameters fixed except the amplitude. A second purpose of the fit was to visually validate the model correspondence to data.

To quantify the spread in threshold levels, the pulse height spectrum of each individual channel was fitted to the mean using amplitude and a translation in threshold level as free parameters. The translation represents the residual from trimming, and was assumed to increase linearly as a function of mean threshold level with a minimum at the trimming point. From the residuals of the individual channels, the channel standard deviation could be calculated, which hence also increases linearly from the trimming point. Standard deviations calculated from several spectra acquired with different kVp were combined with weights provided by statistical errors. When measuring on the low-energy threshold, the high-energy threshold was set to its maximum value so that it would not influence the measurement, and the sum of the high- and low-energy bins was recorded. When measuring on the high-energy threshold, the low-energy threshold was set to approximately half the acceleration voltage to reject virtually all charge shared events but still detect most of the spectrum.

Leakage of the AC logic associated with charge sharing affects image noise, and can be measured with the noise power spectrum (NPS). If a fraction $\chi$ photons are double counted in each channel, three uncorrelated processes can be identified, namely, single counting of the photon with a probability $(1 - \chi)$, or double counting in the right or left adjacent channel with probabilities $\chi/2$ each. In our case, the latter two are equivalent, and for a large number of photons, the autocovariance in the detector direction of the image is, $K(x) = (1 - \chi)K_s(x) + \chi K_d(x)$, where $K_s$ and $K_d$ are the autocovariance functions for single and double counting. For single counting, the image function is a Dirac function ($\delta$), and so is the autocovariance [28], i.e. $K_s(x) = \sigma^2 \delta(x)$, where $\sigma^2$ is the variance. If the quanta are poisson distributed, $\sigma^2 = G^2 \overline{N}$, where $\overline{N}$ is the expectation value of the true number of counts without double counting, and $G$ is the large area gain of the system. In the case of double counting, the image function is instead represented by two Dirac functions, separated by the strip pitch ($p$), and the autocovariance is hence $K_d = G^2 \overline{N}[2\delta(x) + \delta(x-p) + \delta(x+p)]$. The expectation value of number of detected counts in the channel, including double counting, is $\overline{n} = G\overline{N}(1 + \chi)$. Combining the above, and since the NPS of a stationary system is the Fourier transform of the autocovariance [28],

$$
\begin{aligned}
\frac{S(u)}{\overline{n}} &= \frac{(1-\chi)\widehat{K_s}(x) + \chi \widehat{K_d}(x)}{G\overline{N}(1+\chi)} = \\
&= G\frac{1 + \chi[1 + 2\cos(2\pi u/p)]}{1+\chi}, \quad (4)
\end{aligned}
$$

where $S$ is the NPS, $u$ is the spatial frequency in the $x$-direction, and Fourier transforms are denoted by the circumflex. In particular, $S(0) = G(1 + 3\chi)/(1 + \chi)$ when normalized with the mean channel signal, which has been derived previously for $G = 1$ [17].

---

[3]type 23344 and electrometer Unidose E, PTW, Freiburg, Germany

The NPS was measured and calculated in a way similar to standardized methodology as applied to the MDM geometry [16]. 1000 $100 \times 100$ pixel regions of interest (ROI's) were acquired from a flat-field image of 0.5 mm aluminum and 40 mm polymethyl methacrylate at 28 kVp. The NPS was then calculated as the mean of the squared fast Fourier transform of the difference in image signal from the mean in each ROI. $\chi$ was determined from Eq. 4 with the mean ROI signal as $\overline{n}$. In case the NPS in the high-energy bin and chance coincidence is negligible, $\chi = \xi_{cs}$.

The flux in the MDM setup was limited due to technical constraints, and to measure the detector linearity, a similar setup but with a single 128-channel detector module was used. A tungsten target x-ray tube[4] at 33 kVp was filtered with 0.5 mm aluminum, and the flux was controlled with the anode current and an adjustable slit in front of the detector. Levels of the low-energy threshold in individual channels were again trimmed towards the electronic noise, and the global threshold level was set relatively high to reject all noise and most charge-shared events, whereas the global high-energy threshold was set to the maximum value to detect AC events exclusively. The mean of all channels as a function of flux was recorded for both energy bins, with and without AC. In the former case, non-linearity is introduced by pile-up and chance coincidence, but without AC, pile-up only contributes. $\tau_{pu}$, $\tau_{ac}$, and $\xi_{cc}$ were found from Eqs. (1) and (3).

Error estimates of the measurements described above were calculated from the scatter of several data points around the fitted curve assuming a normal distribution, as propagated statistical errors, or as the maximum of these two in case both were available [29]. The estimates are in all cases presented as $\pm 1$ standard deviation. Fitting to measured data was done in a least-squares sense, weighted with propagated statistical errors where applicable.

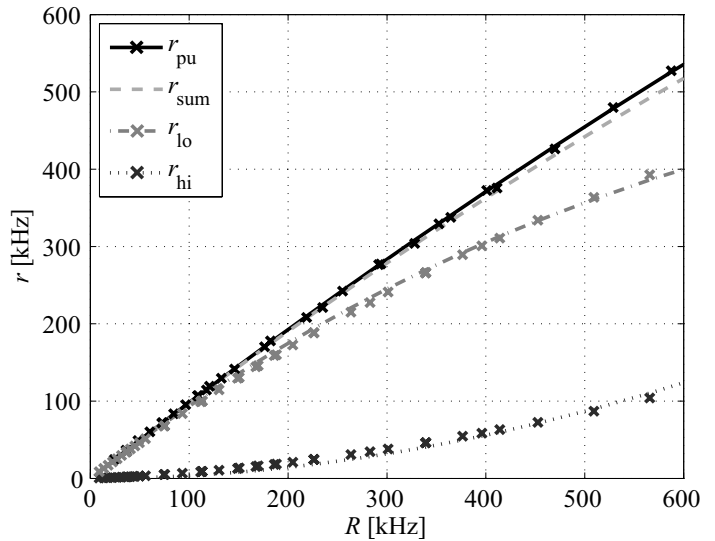

Figure 2: Linearity of the detector as a function of true count-rate ($R$). $r_{pu}$ is the count-rate without anti-coincidence logic. $r_{lo}$, $r_{hi}$, and $r_{sum}$ are count-rates with anti-coincidence logic in the high- and low-energy bins, and the sum of the two. Measurement points are indicated by crosses, and fits to these by Eqs. (1) and (3) are shown with lines.

## 3. Results and Discussion

Figure 2 shows the linearity measurement, with fits to Eqs. (1) and (3) for $r_{pu}$, $r_{lo}$, $r_{hi}$, and $r_{sum} = r_{lo} + r_{hi}$. $R$ was extrapolated from the approximately linear curve through points at low count-rates. The shaper dead time was found to be $\tau_s = 189 \pm 2$ ns, and the AC time window and chance coincidence leakage were $\tau_{ac} = 138 \pm 0.3$ ns and $\xi_{cc} = 0.87 \pm 0.01$, respectively. In all cases, the error estimates from the scatter of the data correspond closely to what is expected from the counting statistics, indicating that the errors are primarily random. $r_{pu}$ and $r_{sum}$ almost coincide, which illustrates that the high leakage results in only a small loss of counts to chance coincidence.

The NPS divided by the mean ROI signal is shown in Fig. 3 for both energy bins. The flux was 33 kHz, which is low enough for pile-up and chance coincidence to be negligible (Fig. 2). Double counting in the high-energy bin results in a bent NPS in the detector direction, and by fitting to Eq. 4, the leakage of the AC logic was found to be $\xi_{cs} = 0.20 \pm 0.002$. The error estimates from the scatter of the data points are small and correspond closely to expectations from statistics, which suggests that Eq. 4 describes the data well. A flat NPS indicates uncorrelated pixels, which, as expected, is the case for the low-energy bin in the detector direction and for both bins in the scan direction.

Figure 4 shows an example of a threshold scan of the high-energy threshold over a 25-kVp spectrum. The scan is shown as a function of global threshold level, which is related to photon energy through Eq. (2), and the cross section at $T_{hi} = 50$ is shown as a histogram to the right.

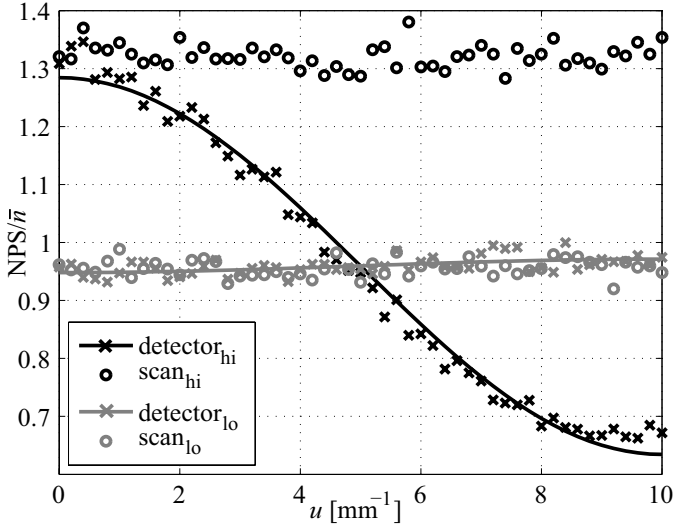

Figure 3: NPS as a function of spatial frequency ($u$) divided by the mean ROI signal ($S/\overline{n}$) of the high- and low-energy bins in the slit and scan directions. Fitting to Eq. (4) is shown with a solid line.

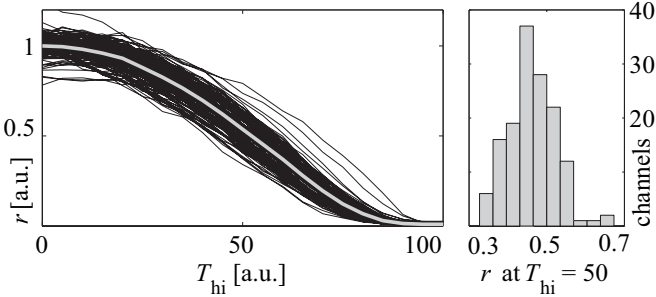

Figure 4: Example of a threshold scan for the high-energy threshold and a 25-kVp spectrum. The thin lines are individual channels, and the mean is indicated in the center. The cross section at $T_{hi} = 50$ is shown to the right.

There is a vertical spread in amplitudes, and a horizontal spread in threshold levels. The former can be compensated for in an image by flat-field calibration, but the threshold spread inevitable reduces energy resolution. Also shown in Fig. 4 is the mean of all channels, which was used as expectation value when estimating the threshold spread and for fitting the model.

Scans of the low-energy threshold are shown in Fig. 5 for the high-energy bin ($r_{hi}$) and for both bins summed ($r_{sum}$). Measurement points are approximately twice as dense as indicated. The high-energy threshold was at its maximum value so that $r_{hi}$ contains AC events exclusively, and the increase towards lower threshold levels is due to increased detection of charge-shared events and increased chance coincidence. Fitting of the model to scans of four spectra in the range 20–40 kVp yielded estimates of the electronic noise and the coefficients ($C$) of Eq. (2). A high flux of $\leq 460$ kHz (decreasing with kVp) was used to cause some amount of pile-up and chance coincidence to challenge the model. The fit is shown in Fig. 5 for three of the spectra, and the global threshold level as a function of energy is superimposed on the figure. The electronic noise was found to be $\sigma_{add} = 4.4$ keV FWHM (505 electrons r.m.s.). Using the relationship of Eq. (2), the threshold spread was translated from threshold levels into $\sigma_{lo} = 2.4 \pm 0.2$ to $2.9 \pm 0.2$ keV FWHM in the interval 1–20 keV, which is where the low-energy threshold is supposed to operate. Error estimates were propagated from the statistical uncertainty of the threshold spread. In units of threshold levels, the spread was found to be fairly constant with global threshold level, and the increase towards higher energies is mainly due to the nonlinearity of Eq. (2).

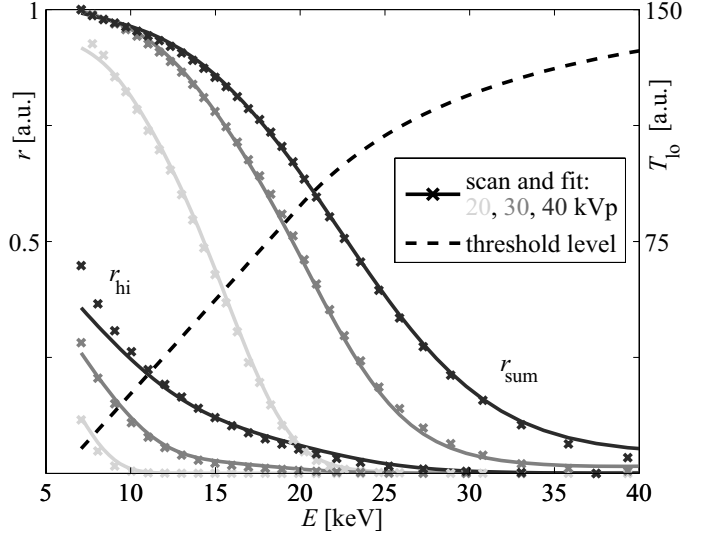

Figure 5: Scans of the low-energy threshold over 20–40 kVp spectra with the high-energy threshold at its maximum value. Two groups of curves are shown corresponding to counts in the low- and high-energy bins summed ($r_{sum}$), and in the high-energy bin only ($r_{hi}$). The low-energy threshold level as a function of energy is shown as a dashed line.

Scans of the high-energy threshold are shown in Fig. 6 for $r_{hi}$. $r_{sum}$ can in this case be assumed constant and is therefore not shown. All measurement points are indicated. The low-energy thresholds were set to 10.7, 12.3, 18.9, 20.6, and 22.6 keV for the five 20–40 kVp spectra, with levels and spread determined by the low-energy threshold scan. A constant background is evident for scans above 30 kVp, which is due mainly to chance coincidence and not charge sharing because the low-energy thresholds were relatively high. The electronic noise and the coefficients of Eq. 2 were fitted, with the resulting model prediction and relationship between threshold and energy shown in Fig. 6. $\sigma_{add}$ was 2.9 keV FWHM (339 electrons r.m.s.). The spread of the thresholds was assumed to be $\sigma_{hi} = 0$ at 33.2 keV. Below the trimming point, the spread found a maximum of $\sigma_{hi} = 1.2 \pm 0.4$ keV FWHM at 20 keV, and it increased rapidly and monotonically above the trimming point, reaching $\sigma_{hi} = 1.9 \pm 0.7$ keV at 40 keV. Again, the spread towards higher energies is strongly enhanced by the nonlinearity of $T(E)$.

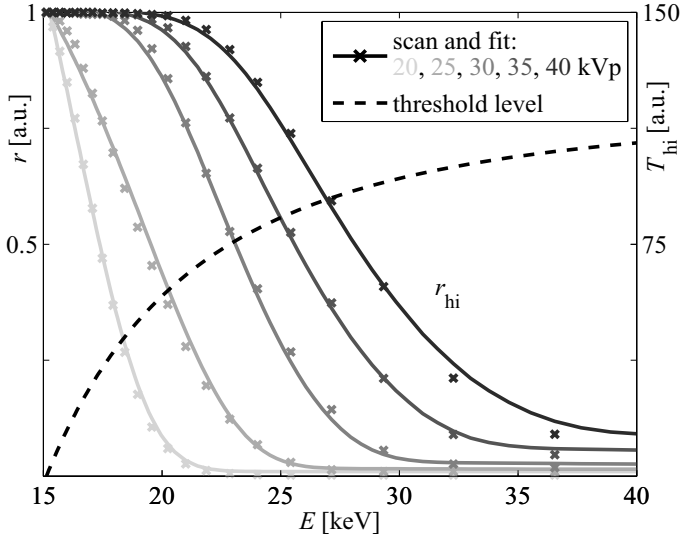

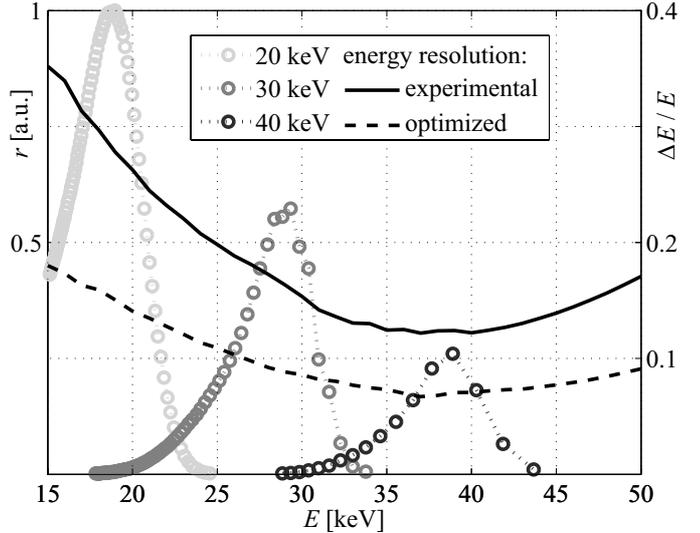

Figure 6: Scans of the high-energy threshold measured in the high-energy bin ($r_{hi}$) over 20–40 kVp spectra with low-energy thresholds at approximately half the acceleration voltage. The high-energy threshold level as a function of energy is shown as a dashed line.

Figure 7: Energy response on monochromatic delta peaks. The plotted peaks are for the experimental detector. Energy resolution ($\Delta E/E$) is shown for the experimental detector, and for an improved detector with high AC efficiency and low threshold spread and electronic noise.

As a general observation it can be said that the model agrees reasonably well with measured data. Statistical errors of the scans in Figs. 5 and 6 are small, and it is clear that systematic errors, caused by assumptions in the model and errors in all input parameters, dominate for the fit. Valid error estimates on $C$ and $\sigma_{add}$ are thus hard to obtain.

Figure 7 illustrates the energy response of the high-energy threshold to delta peaks at low count-rates (no pile-up or chance coincidence). The experimental detector was evaluated with the low-energy threshold at 7 keV. Response functions at 20, 30, and 40 keV are plotted with $\Delta E = 5.3$, 4.6, and 4.9 keV. As expected, peak widths increase away from the trimming point because of increased threshold spread, whereas charge sharing spreads all peaks towards lower energies. Simulation points are plotted in steps of the maximum bit depth of the threshold, and the increasing spread at higher energies reflects the contribution to peak width caused by the nonlinear shaper output. In fact, one step in threshold level corresponds to 1.4 keV at 40 keV, but only 0.13 keV at 20 keV. The peak heights correspond to the relative amount of energy resolved information, and the decline towards higher energies is due to decreased quantum efficiency and increased detection of charge-shared events that results in a constant background in the high-energy bin.

$\Delta E/E$ is plotted in Fig. 7 as a function of energy. For the 20, 30, and 40 keV peaks, $\Delta E/E = 0.26$, 0.15, and 0.12 respectively. Assuming that the largest source of random errors was the threshold spread, propagated relative errors on the energy resolution were less than 4.5%, and it is hence likely that systematic errors dominate. In summary, the largest contributors to the peak broadening are the electronic noise (2.9 keV FWHM), followed by threshold spread (1.2–1.9 keV FWHM), and charge sharing (1.8–1.4 keV FWHM). Note that our particular choice of FWHM as $\Delta E$ measure slightly underestimates the contribution by charge sharing because peak tails are neglected.

The energy resolution of the present detector is lower than some previously reported results on similar silicon strip detectors [14], which is, however, due mainly to the fact that we have considered a full system in this study. For instance, the small strip pitch needed for high-resolution mammography causes relatively large amounts of electronic noise and charge sharing, the double-threshold configuration adds complexity and electronic noise, and channel-to-channel threshold spread reduces the energy resolution when more than one channel is considered. The predicted energy resolution of an improved detector with half the threshold spread and electronic noise, and with a 3.5-keV low-energy threshold and no leakage of the AC logic is also shown in Fig. 7 for comparison. Improvements of 1.9–1.7 times are seen at 20–40 keV. Note that this is still substantially worse than the intrinsic energy resolution of silicon, which is in the order of 0.01 in the interval.

One aspect of energy resolution that is not captured by the $\Delta E/E$ measure is the constant background of AC events that are put in the high-energy bin. For the experimental detector in Fig. 7, charge sharing results in a background intensity of 24–63% for the three peaks. Lowering the low-energy threshold, as for the near-ideal case in Fig. 7, results in more efficient AC and a narrower peak, but also more background without energy information. At high intensities, pile-up and chance coincidence would also contribute to a more or less constant background.

## 4. Conclusions

Measurements, simulations, and published data were used as input parameters to a cascaded detector model, which was validated by comparison to threshold scans over several input spectra. Using the model, the energy response of the detector assembly could be assessed on a system level without monochromatic radiation, and the impact of various parameters could be estimated.

The energy resolution was found to be $\Delta E/E = 0.12$–$0.26$ in the relevant energy range. The major factors contributing to the width of the response function were found to be electronic noise, followed by charge sharing, and a channel-to-channel threshold spread that was boosted by a nonlinear shaper output. Additionally, a relatively large constant background of charge-shared photons detected by the AC logic was added to the high-energy bin. The shaper dead time and AC time window were both less than 200 ns, and pile-up and chance coincidence were found to be of minor importance at mammography count-rates. Fluorescence and scattering effects in the silicon were estimated to be negligible.

Relatively large improvements of the energy resolution are within reach. Minimization of the electronic noise is highly important to reduce the peak broadening. The trimming point should be chosen close to the point of operation of the threshold, and variations between channels should be kept at a minimum in order to minimize the threshold spread. This is particularly important for the high-energy threshold, which is meant to operate in high-intensity parts of the spectrum. An improvement in shaper and discriminator linearity at higher energies is also desirable to reduce the effects of threshold spread and limited bit depth. Finally, the AC scheme can be improved by keeping the energy resolution of detected events, or by recording them in a separate bin.

Preliminary studies already indicate clinical benefit for spectral imaging with the described detector [30]. The information and model provided here will be crucial for the ongoing system optimization.

## 5. Acknowledgments

The authors wish to thank Magnus Hemmendorff and Alexander Chuntonov at Sectra Mamea AB for discussions and practical help with measurements, and Björn Svensson, also at Sectra, for discussions on charge sharing. This work was funded in part by the European Union through the HighReX project.